\documentclass[conference,a4paper]{IEEEtran}

\usepackage{cite}
\usepackage{amsmath,amssymb,amsfonts}
\usepackage{graphicx}
\usepackage[table,dvipsnames]{xcolor}
\usepackage{booktabs}
\usepackage{multirow}
\usepackage{array}
\usepackage{siunitx}
\usepackage{placeins}
\usepackage{float}
\usepackage{tikz}
\usepackage{pgfplots}
\pgfplotsset{compat=1.18}
\usepackage{hyperref}
\usepackage{xcolor}
\usepackage{tabularx}
\usetikzlibrary{shadows}

\hypersetup{
    colorlinks=true,      
    linkcolor=blue,       
    filecolor=magenta,    
    urlcolor=blue,        
    pdftitle={Overleaf Example},
}

\newcommand{\na}{--}

\newcolumntype{L}[1]{>{\raggedright\arraybackslash}p{#1}}

\definecolor{MerconBlue}{RGB}{35,78,122}
\definecolor{MerconTeal}{RGB}{26,138,140}
\definecolor{MerconGold}{RGB}{214,161,48}
\definecolor{MerconCoral}{RGB}{201,93,65}
\definecolor{MerconSlate}{RGB}{80,89,102}

\definecolor{TableGold}{RGB}{255,215,0}      
\definecolor{TableSilver}{RGB}{192,192,192}  
\definecolor{TableBronze}{RGB}{205,127,50}   

\definecolor{CellGold}{RGB}{255,240,170}     
\definecolor{CellSilver}{RGB}{225,225,230}   
\definecolor{CellBronze}{RGB}{238,210,175}   

\newcommand{\ColHeader}[1]{\textbf{#1}}

\def\BibTeX{{\rm B\kern-.05em{\sc i\kern-.025em b}\kern-.08em
    T\kern-.1667em\lower.7ex\hbox{E}\kern-.125emX}}

\begin{document}

\title{A Controlled Visual-Backbone Benchmark for\\
       Multimodal Short-Term Solar Irradiance Forecasting}

\author{%
\IEEEauthorblockN{%
  \parbox[t]{0.22\textwidth}{\centering
    Oshadha Samarakoon\\
    {\small\textit{Dept.\ of EEE, University of Peradeniya}\\
    Peradeniya, Sri Lanka\\
    e21345@eng.pdn.ac.lk}
  }\hfill
  \parbox[t]{0.22\textwidth}{\centering
    Dushan Herath\\
    {\small\textit{MARC, University of Peradeniya}\\
    Peradeniya, Sri Lanka\\
    dushan.herath@eng.pdn.ac.lk}
  }\hfill
  \parbox[t]{0.22\textwidth}{\centering
    Ishara Ranmandala\\
    {\small\textit{MARC, University of Peradeniya}\\
    Peradeniya, Sri Lanka\\
    ranmandalaishara@gmail.com}
  }\hfill
  \parbox[t]{0.22\textwidth}{\centering
    Dilshara Herath\\
    {\small\textit{MARC, University of Peradeniya}\\
    Peradeniya, Sri Lanka\\
    dilshara.herath3@gmail.com}
  }\\[10pt]
  \makebox[\textwidth][c]{%
    \parbox[t]{0.28\textwidth}{\centering
      Roshan Godaliyadda\\
      {\small\textit{Dept.\ of EEE, University of Peradeniya}\\
      Peradeniya, Sri Lanka\\
      roshang@eng.pdn.ac.lk}
    }\hspace{0.04\textwidth}
    \parbox[t]{0.28\textwidth}{\centering
      Parakrama Ekanayake\\
      {\small\textit{Dept.\ of EEE, University of Peradeniya}\\
      Peradeniya, Sri Lanka\\
      mpbe@eng.pdn.ac.lk}
    }\hspace{0.04\textwidth}
    \parbox[t]{0.28\textwidth}{\centering
      Vijitha Herath\\
      {\small\textit{Dept.\ of EEE, University of Peradeniya}\\
      Peradeniya, Sri Lanka\\
      vijitha@eng.pdn.ac.lk}
    }%
  }%
}}

\maketitle

\begin{abstract}
Sky-image irradiance studies often compare forecasting systems in which the image encoder, temporal model, fusion block, target definition, and training recipe all change together. We use a narrower protocol: the multimodal forecasting pipeline is fixed, and only the visual backbone is varied. The shared setup keeps preprocessing, clear-sky-index normalization, weather-history encoding, fusion, regression head, loss, optimizer schedule, seed, and chronological split policy unchanged. We compare ConvNeXt, Swin Transformer, VMamba, Spatial Mamba, and MambaVision backbones for \SI{10}{min}-ahead forecasting on Folsom and a strict matched NREL split. Forecast skill is measured against clear-sky-index smart persistence, and temporal-only rows are reported as weather-history diagnostics rather than as the main ranking criterion. On the Folsom strict split, all evaluated visual-backbone runs improve over smart persistence. In the evaluated single-seed strict runs, VMamba Small and Swin Base reach matched Folsom RMSE values of \SI{65.39}{\watt\per\metre\squared} and \SI{65.50}{\watt\per\metre\squared}; the temporal-only diagnostic reaches \SI{69.51}{\watt\per\metre\squared}. On the \num{313}-sample NREL strict split, smart persistence remains strongest at \SI{17.48}{\watt\per\metre\squared}, while the lowest visual RMSE is obtained by Swin Tiny at \SI{23.76}{\watt\per\metre\squared}. These results provide a reproducible encoder comparison under one fixed multimodal operating point rather than establishing architecture-level dominance, statistically resolved ranking, or fully optimized forecasting performance.Code available here: \url{https://github.com/Oshadha345/irradiance_benchmark}
\end{abstract}


\section{Introduction}

Very-short-term solar irradiance forecasting is important because
photovoltaic output can change within minutes. Ground-based sky
images capture local cloud conditions, while recent irradiance and
meteorological histories describe the current operating state.
However, many image-based forecasting studies vary the visual
encoder, temporal model, fusion strategy, target definition, and
optimizer together, making it difficult to isolate the effect of
the visual backbone.

The encoder choice is now broad. ConvNeXt modernizes convolutional
networks \cite{liu2022convnet2020s}, Swin Transformer uses
hierarchical shifted-window attention
\cite{liu2021swintransformerhierarchicalvision}, and visual
state-space models such as VMamba, Spatial Mamba, and MambaVision
adapt selective state-space ideas to image feature maps
\cite{liu2024vmambavisualstatespace,xiao2025spatialmambaeffectivevisualstate,hatamizadeh2025mambavisionhybridmambatransformervision}.
These families differ in capacity, spatial aggregation, and
throughput, but such differences are difficult to interpret when
each encoder is evaluated inside a different forecasting pipeline.

This paper uses a narrower protocol. Rather than tuning a separate
system for each backbone, we fix the temporal branch, projection
layers, fusion rule, target definition, optimizer, seed, and
chronological split policy, and vary only the visual backbone. The
shared architecture uses a four-stage visual projector, a
single-layer LSTM over a \num{40}-step weather history, direct
concatenation, and a lightweight regression head; the experiment
therefore measures how pretrained visual encoders behave inside the
same \SI{10}{min}-ahead multimodal irradiance forecaster.

\section{Related Work}

Ground-based sky-image forecasting uses local cloud observations
for short lead times \cite{NIE2024113977}. Classical methods used
cloud tracking, optical flow, shadow maps, and clear-sky models
\cite{CALDAS20191643}, while recent deep learning methods use CNN,
MLP, LSTM, or Transformer modules
\cite{ELALANI2021888,HENDRIKX2024112463,LIU2023121160}.
Multi-location studies also show that camera setup, sample count,
and local weather regime can strongly affect reported forecasting
performance \cite{NIE2024113977,NIE2024123467}.

Multimodal irradiance forecasting combines sky images with GHI or
weather history, but the temporal branch can already carry much of
the short-horizon signal during stable conditions. Clear-sky-index
normalization reduces the deterministic solar-geometry trend
\cite{solar2040026}, making persistence and smart persistence
necessary operational baselines
\cite{CALDAS20191643,STRAUB2024112319}. Image-based clear-sky-index
methods can improve forecasts under variable skies, although abrupt
transitions and overcast regimes remain difficult
\cite{MARTINEZLOPEZ2024112320,Varaschin_2025}.

Visual encoder families now include modern CNNs, hierarchical
Transformers, and visual state-space models. Mamba introduced
selective state spaces for linear-time sequence modeling
\cite{gu2024mambalineartimesequencemodeling}; related visual
variants adapt these ideas to image feature maps through spatial
scanning or hybrid attention designs
\cite{liu2024vmambavisualstatespace,xiao2025spatialmambaeffectivevisualstate,hatamizadeh2025mambavisionhybridmambatransformervision}.
Controlled backbone benchmarking across this diverse family has
recently shown value in remote-sensing segmentation, where
domain-shift and boundary sensitivity vary substantially across
CNN, Transformer, and SSM encoders under a fixed evaluation
contract \cite{vsscd}.
This motivates an analogous controlled comparison for sky-image
irradiance forecasting, in which the non-visual components are
fixed and the image encoder is isolated as the varied component.

The paper makes three contributions:
\begin{enumerate}
    \item a visual-backbone comparison protocol for multimodal
          sky-image irradiance forecasting, with preprocessing,
          clear-sky-index target definition, temporal encoding,
          fusion, regression head, optimizer schedule, seed, and
          chronological split fixed;
    \item an evaluation of CNN, Transformer, and visual state-space
          backbones using clear-sky-index smart persistence as the
          operational skill baseline and temporal-only rows as
          weather-history diagnostics;
    \item accuracy, bias, forecast skill, throughput, model size,
          FLOPs, Pareto behavior, and qualitative ERF overlays for
          forecasting and deployment-oriented interpretation.
\end{enumerate}

The remainder of the paper is organized as follows. Section III
describes the methodology, Section IV, experimental setup,
Section V, results and discussion, and Section VI concludes.
\section{Methodology}

\subsection{Forecasting Task}

Let $I_t \in \mathbb{R}^{H \times W \times 3}$ denote the masked sky image at time $t$, and let $X_t \in \mathbb{R}^{40 \times 7}$ denote the preceding weather-history sequence. The model predicts the \SI{10}{min}-ahead clear-sky index,
\begin{equation}
\hat{k}_{t+10} = f(I_t, X_t), \qquad
k_{t+10} = \frac{\mathrm{GHI}_{t+10}}{\mathrm{GHI}^{\mathrm{cs}}_{t+10}},
\end{equation}
where $\mathrm{GHI}^{\mathrm{cs}}$ is computed with the Ineichen clear-sky model. Samples with missing values are removed, night and low-sun samples with solar zenith angle above \SI{85}{\degree} are excluded, and $k$ is clipped to $[0,1.2]$ before standardization. The clear-sky index and meteorological features are standardized using statistics fitted only on the chronological train/validation window. At evaluation, predictions and targets are inverse-transformed to $k$ and multiplied by $\mathrm{GHI}^{\mathrm{cs}}_{t+10}$, so errors are reported in \si{\watt\per\metre\squared}.

\subsection{Fixed Forecasting Architecture}

\begin{figure*}[!t]
\centering
\includegraphics[height=0.40\textheight,keepaspectratio]{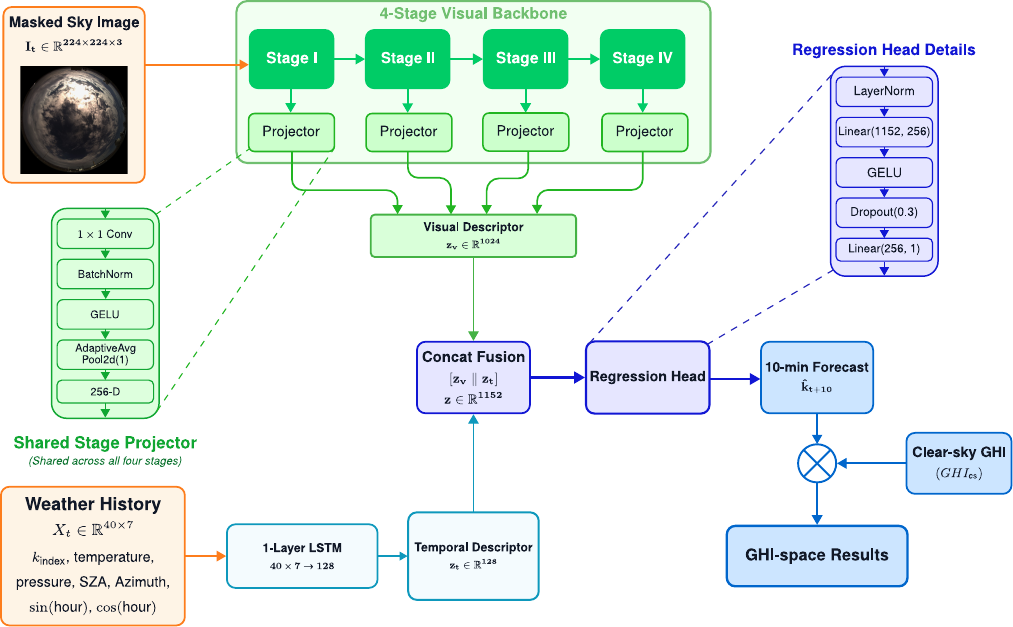}
\caption{Shared multimodal forecasting architecture. The masked sky image and weather-history sequence are encoded into $z_v$ and $z_t$, concatenated, and passed through the same regression head for all runs; only the visual backbone family and scale are varied.}
\label{fig:architecture}
\end{figure*}

Fig.~\ref{fig:architecture} shows the shared multimodal architecture. Images are resized to \num{224}$\times$\num{224}; a centered circular sky-dome mask with radius $250/512$ of the image width is applied after resizing, and pixels outside the dome are set to zero before ImageNet normalization.

Each visual backbone exposes four feature stages $\{F_i\}_{i=1}^{4}$. All five families produce hierarchical four-stage feature pyramids by design: ConvNeXt and Swin expose these through the \texttt{timm} feature-extractor interface, while VMamba, Spatial Mamba, and MambaVision expose them through their respective repository implementations. Stage-specific projectors match the incoming channels but use the same form: a \num{1}$\times$\num{1} convolution, batch normalization, GELU, and global average pooling. Each projector outputs \num{256} channels, giving
\begin{equation}
z_v =
\mathrm{concat}\!\left[
\mathrm{GAP}(\phi_1(F_1)),\ldots,
\mathrm{GAP}(\phi_4(F_4))
\right]
\in \mathbb{R}^{1024}.
\end{equation}
Thus, every backbone passes the same visual-descriptor dimension to the regression head.

The temporal branch is a single-layer LSTM over seven channels: clear-sky index, temperature, pressure, solar zenith angle, solar azimuth, and sine/cosine time-of-day encodings. It returns
\begin{equation}
z_t = \mathrm{LSTM}(X_t) \in \mathbb{R}^{128}.
\end{equation}
The fused descriptor $z=[z_v \,\|\, z_t] \in \mathbb{R}^{1152}$ is mapped to a scalar by layer normalization, a \num{256}-unit GELU layer, dropout with rate \num{0.3}, and a final linear layer. Training uses Huber loss with $\delta=1.0$.

\subsection{Backbones and Weather-History Diagnostic}

ConvNeXt and Swin are loaded through the \texttt{timm} feature-extractor interface, while VMamba, Spatial Mamba, and MambaVision use repository implementations with ImageNet-pretrained weights. The full architecture---backbone, stage projectors, LSTM branch, and regression head---is trained end-to-end in a single pass; backbone parameters are fine-tuned at \num{0.1}$\times$ the task-head learning rate. These backbones are evaluated under the fixed protocol described in Section~\ref{sec:experimental_setup}.

The temporal-only diagnostic replaces $z_v$ with a zero vector of the same dimension, preserving the fusion interface while keeping the LSTM, regression head, loss, optimizer, and data split unchanged. This row measures the weather-history branch inside the fixed architecture; it is not a separately optimized temporal model and is not used to rank visual backbones.

\subsection{Smart Persistence Baseline}

Smart persistence is computed in clear-sky-index space:
\begin{equation}
\hat{k}^{\mathrm{sp}}_{t+10}=k_t,\qquad
\widehat{\mathrm{GHI}}^{\mathrm{sp}}_{t+10}
= k_t\,\mathrm{GHI}^{\mathrm{cs}}_{t+10}.
\end{equation}
Here $k_t$ is the most recent clear-sky index in the \num{40}-step weather history. The baseline is evaluated on the same retained samples as each model and is used as the forecast-skill denominator.

All learned models, the temporal-only diagnostic, and smart persistence are evaluated under the fixed comparison protocol described next.
\section{Experimental Setup}
\label{sec:experimental_setup}

\subsection{Datasets and Splits}

We evaluate on the Folsom and NREL sky-image datasets using
chronological splits to avoid leakage: Folsom uses 2014--2015 for
train/validation and 2016 for testing, while NREL uses 2018--2019
for train/validation and 2020 for testing. A sample is retained only
when the image at time $t$, the full \num{40}-step weather history,
and the \SI{10}{min}-ahead target are available; the same matched
samples are used for all visual rows and the temporal-only
diagnostic. Table~\ref{tab:dataset_summary} reports post-filter
loader counts, not raw archive sizes.

\begin{table}[htb]
\centering
\caption{Dataset summary after image matching, history checks,
         and target filtering.}
\label{tab:dataset_summary}
\small
\setlength{\tabcolsep}{5pt}
\renewcommand{\arraystretch}{1.1}
\begin{tabular}{lrrrr}
\toprule
\ColHeader{Dataset} & \ColHeader{Train} & \ColHeader{Val} &
\ColHeader{Test} & \ColHeader{Test year} \\
\midrule
Folsom & 385{,}115 & 47{,}598 & 224{,}022 & 2016 \\
NREL   &       580 &       64 &       313 & 2020 \\
\bottomrule
\end{tabular}
\end{table}

The \num{40}-step history corresponds to about \SI{39}{min} on
Folsom at \SI{60}{s} cadence and about \SI{6.5}{h} on NREL at
\SI{10}{min} cadence. After daylight filtering, image matching, and
history checks, NREL has only \num{313} strict test samples; we
therefore treat it as a low-data matched-sample stress test rather
than a full-archive estimate of NREL forecasting performance.

\begin{table}[!t]
\centering
\caption{Fixed \texttt{strict\_v1} comparison protocol used for
         all evaluated runs.}
\label{tab:benchmark_protocol}
\small
\setlength{\tabcolsep}{6pt}
\renewcommand{\arraystretch}{1.3}
\begin{tabular}{p{0.20\linewidth} p{0.72\linewidth}}
\toprule
\ColHeader{Axis} & \ColHeader{Setting} \\
\midrule
Varied factor
  & ConvNeXt T/S/B/L, Swin T/S/B, VMamba T/S/B,
    Spatial Mamba T/S/B, MambaVision T/T2/S/B/L \\[3pt]
Input processing
  & \num{224}$\times$\num{224} resize, circular sky-dome mask,
    ImageNet normalization \\[3pt]
Target
  & \SI{10}{min}-ahead clear-sky index;
    inverse-transformed to physical GHI units \\[3pt]
Sample construction
  & Matched image, \num{40}-step weather history, future target,
    daylight and low-sun filtering \\[3pt]
Temporal pathway
  & Seven weather-history channels, single-layer LSTM,
    \num{128}-D descriptor \\[3pt]
Visual interface
  & Four stages, fixed-form projectors, \num{256} channels per
    stage, \num{1024}-D descriptor \\[3pt]
Fusion/head
  & Concatenation, layer normalization, \num{256}-unit GELU,
    dropout \num{0.3}, linear output \\[3pt]
Optimization
  & Huber loss, AdamW, batch size \num{32}, \num{8} epochs,
    seed \num{42}, cosine schedule, no early stopping \\[3pt]
Evaluation
  & Chronological split, smart-persistence skill, RMSE, MAE,
    MBE, $R^{2}$, parameters, FLOPs, FPS, ERF \\
\bottomrule
\end{tabular}
\end{table}

\subsection{Fixed Protocol and Metrics}

All runs follow the fixed comparison protocol summarised in
Table~\ref{tab:benchmark_protocol}. The only intentionally varied
factor is the visual backbone family and scale; all data filtering,
target construction, architecture interface, optimization choices,
and evaluation procedures are fixed. This improves comparability
across encoders but does not assume that the shared head is
individually optimal for every backbone.

All learned runs use AdamW with learning rate $5\times10^{-5}$,
backbone learning-rate ratio \num{0.1}, weight decay \num{0.05},
cosine decay to $10^{-6}$, and Python, NumPy, and PyTorch seeds set
to \num{42} before model and data-loader construction. Validation is
used only to select the saved checkpoint within each fixed run; no
model variant is changed after test evaluation.

RMSE is reported in physical GHI units, and forecast skill is
computed relative to smart persistence:
\begin{equation}
  \mathrm{FS}_{\mathrm{sp}}
    = 100\!\left(1 - \frac{\mathrm{RMSE}_{\mathrm{model}}}
                           {\mathrm{RMSE}_{\mathrm{sp}}}\right).
\end{equation}
Efficiency is measured on a single NVIDIA Quadro GV100
\SI{32}{GB} GPU using batch size \num{4}, random inputs,
evaluation mode, \num{10} warmup iterations, and \num{50} timed
iterations. We also report MAE, mean bias error, $R^{2}$,
parameter count, FLOPs, FPS, and qualitative ERF overlays.
\section{Results and Discussion}

\subsection{Accuracy and Efficiency}

Table~\ref{tab:main_results} summarises the evaluated
\texttt{strict\_v1} runs. On Folsom, all evaluated visual-backbone
runs improve over smart persistence. VMamba~S gives the lowest
Folsom visual RMSE, \SI{65.39}{\watt\per\metre\squared}, with
19.64\% forecast skill. Swin~B is nearly tied at
\SI{65.50}{\watt\per\metre\squared}, followed by Spatial Mamba~S at
\SI{65.99}{\watt\per\metre\squared}. The temporal-only diagnostic
reaches \SI{69.51}{\watt\per\metre\squared}; it contextualises the
weather-history branch but is not used to rank visual encoders.

On the \num{313}-sample NREL strict split, the long physical history
window (\SI{6.5}{h} at \SI{10}{min} cadence) strengthens
persistence-like temporal information, while the small matched image
set limits reliable visual learning; smart persistence therefore
remains strongest at \SI{17.48}{\watt\per\metre\squared}. The
temporal-only diagnostic reaches \SI{21.33}{\watt\per\metre\squared},
while the lowest visual RMSE is obtained by Swin~T at
\SI{23.76}{\watt\per\metre\squared}. Since every learned visual row
has negative skill relative to smart persistence, this split is used
to observe behaviour under extreme matched-sample scarcity, not as a
full-site visual-backbone ordering.

VMamba~S gives the lowest Folsom RMSE but runs at 168.5~FPS.
ConvNeXt~T and Swin~T run at 561.5 and 471.5~FPS respectively,
while keeping Folsom RMSE near \SI{66.3}{\watt\per\metre\squared}.
These are full forecasting-model throughput measurements, not
standalone backbone speeds. Throughput is used here as a
computational-cost indicator for deployment-oriented comparison
rather than as a claim that sub-second inference is required for a
\SI{10}{min}-ahead task.

\begin{table*}[!t]
\centering
\caption{Strict-protocol results for evaluated runs.
  RMSE is in \si{\watt\per\metre\squared};
  FS is relative to smart persistence.
  Efficiency is measured with \num{224}$\times$\num{224} inputs
  and batch size~4.
  \colorbox{CellGold}{\strut Gold}, \colorbox{CellSilver}{\strut silver},
  and \colorbox{CellBronze}{\strut bronze} cells mark the top-3 visual
  models per site.
  \textbf{NREL has only 313 matched test samples and is reported as a
  low-data stress test; all visual rows yield negative skill relative
  to smart persistence and should not be read as a primary
  site-level backbone ranking.}}
\label{tab:main_results}
\small
\setlength{\tabcolsep}{4pt}
\renewcommand{\arraystretch}{1.22}

\begin{tabular}{llrrrrrrrr}
\toprule
\ColHeader{Family}
  & \ColHeader{Backbone}
  & \ColHeader{Params (M)}
  & \ColHeader{FLOPs (G)}
  & \ColHeader{FPS}
  & \multicolumn{2}{c}{\ColHeader{Folsom}}
  & \multicolumn{2}{c}{\ColHeader{NREL}} \\
\cmidrule(lr){6-7}\cmidrule(lr){8-9}
& & & & &
  \ColHeader{RMSE} & \ColHeader{FS (\%)} &
  \ColHeader{RMSE} & \ColHeader{FS (\%)} \\
\midrule

Baseline
  & Smart persistence & \na & \na & \na
  & 81.37 & 0.00 & 17.48 & 0.00 \\
  & Temporal-only     & 0.4 & 0.0 & 13708.6
  & 69.51 & 14.57 & 21.33 & $-$22.00 \\
\midrule

CNN
  & ConvNeXt T & 28.6 & 18.5 & 561.5
    & 66.25 & 18.58
    & \cellcolor{CellSilver}27.09 & \cellcolor{CellSilver}$-$54.96 \\
  & ConvNeXt S & 50.2 & 35.4 & 338.7
    & 66.49 & 18.28 & 29.14 & $-$66.69 \\
  & ConvNeXt B & 88.4 & 62.3 & 220.9
    & 66.85 & 17.84 & 36.43 & $-$108.39 \\
  & ConvNeXt L & 197.3 & 138.8 & 125.3
    & 67.03 & 17.62 & 35.77 & $-$104.58 \\
\midrule

Transformer
  & Swin T & 28.3 & 18.6 & 471.5
    & 66.45 & 18.33
    & \cellcolor{CellGold}23.76 & \cellcolor{CellGold}$-$35.94 \\
  & Swin S & 49.6 & 35.7 & 258.7
    & 66.52 & 18.25 & 35.56 & $-$103.41 \\
  & Swin B & 87.6 & 62.7 & 195.5
    & \cellcolor{CellSilver}65.50 & \cellcolor{CellSilver}19.50
    & 31.54 & $-$80.41 \\
\midrule

SSM
  & VMamba T & 30.2 & 20.0 & 222.9
    & 66.91 & 17.76 & 33.01 & $-$88.82 \\
  & VMamba S & 50.1 & 34.8 & 168.5
    & \cellcolor{CellGold}65.39 & \cellcolor{CellGold}19.64
    & 33.99 & $-$94.40 \\
  & VMamba B & 88.4 & 61.3 & 135.9
    & 66.21 & 18.63 & 34.02 & $-$94.63 \\
\cmidrule(lr){2-9}

  & Spatial Mamba T & 26.9 & 17.7 & 187.5
    & 68.07 & 16.34 & 39.64 & $-$126.73 \\
  & Spatial Mamba S & 43.2 & 28.0 & 89.3
    & \cellcolor{CellBronze}65.99 & \cellcolor{CellBronze}18.89
    & 35.71 & $-$104.25 \\
  & Spatial Mamba B & 95.7 & 62.2 & 39.6
    & 66.39 & 18.41 & 38.07 & $-$117.75 \\
\cmidrule(lr){2-9}

  & MambaVision T  & 32.6  &  18.1 & 420.3
    & 67.26 & 17.33 & 33.26 & $-$90.27 \\
  & MambaVision T2 & 35.9  &  20.7 & 236.3
    & 66.04 & 18.83 & 35.86 & $-$105.12 \\
  & MambaVision S  & 51.1  &  30.3 & 395.8
    & 67.91 & 16.54 & 34.32 & $-$96.30 \\
  & MambaVision B  & 98.8  &  60.3 & 256.5
    & 66.48 & 18.30 & 32.21 & $-$84.26 \\
  & MambaVision L  & 229.4 & 140.1 & 127.8
    & 66.56 & 18.19
    & \cellcolor{CellBronze}28.91 & \cellcolor{CellBronze}$-$65.35 \\
\bottomrule
\end{tabular}
\end{table*}

\subsection{Pareto and Secondary Metrics}

\begin{figure}[t]
  \centering
  \begin{tikzpicture}
\begin{axis}[
    width=\columnwidth,
    height=0.78\columnwidth,
    xmode=log,
    log basis x=10,
    xmin=35,
    xmax=680,
    ymin=65.1,
    ymax=68.4,
    xlabel={Throughput (FPS, higher is better)},
    ylabel={RMSE (\si{\watt\per\metre\squared}, lower is better)},
    xtick={40,60,100,150,200,300,450,600},
    xticklabels={40,60,100,150,200,300,450,600},
    grid=both,
    major grid style={draw=MerconSlate!25},
    minor grid style={draw=MerconSlate!12},
    legend style={
        draw=MerconSlate!40,
        fill=white,
        fill opacity=0.92,
        font=\small,                    
        legend columns=3,
        at={(0.5,-0.24)},
        anchor=north,
        column sep=10pt,
        row sep=2pt,
        inner sep=8pt,
        rounded corners=8pt,            
        drop shadow={shadow xshift=0.5pt, shadow yshift=-0.5pt},
    },
    tick label style={font=\scriptsize},
    label style={font=\footnotesize},
    every axis plot/.append style={line width=1.1pt},
    clip=false
]

\addplot[only marks, color=MerconBlue, mark=square*, mark size=3pt] 
    coordinates {(561.5,66.251) (338.7,66.492) (220.9,66.847) (125.3,67.033)};
\addlegendentry{ConvNeXt}

\addplot[only marks, color=MerconTeal, mark=triangle*, mark size=3.2pt] 
    coordinates {(471.5,66.453) (258.7,66.519) (195.5,65.501)};
\addlegendentry{Swin}

\addplot[only marks, color=MerconGold!90!black, mark=*, mark size=3pt] 
    coordinates {(222.9,66.913) (168.5,65.388) (135.9,66.206)};
\addlegendentry{VMamba}

\addplot[only marks, color=MerconCoral, mark=diamond*, mark size=3.1pt] 
    coordinates {(187.5,68.069) (89.3,65.994) (39.6,66.387)};
\addlegendentry{Spatial Mamba}

\addplot[only marks, color=gray, mark=pentagon*, mark size=3.2pt] 
    coordinates {(420.3,67.264) (236.3,66.043) (395.8,67.910) (256.5,66.479) (127.8,66.564)};
\addlegendentry{MambaVision}

\node[anchor=east, font=\scriptsize\bfseries, text=MerconBlue, fill=white, fill opacity=0.92, inner sep=1.5pt, rounded corners=2pt] at (axis cs:561.5,66.251) {T};
\node[anchor=north west, font=\scriptsize\bfseries, text=MerconTeal, fill=white, fill opacity=0.92, inner sep=1.5pt, rounded corners=2pt] at (axis cs:195.5,65.501) {B};
\node[anchor=south west, font=\scriptsize\bfseries, text=MerconGold!90!black, fill=white, fill opacity=0.92, inner sep=1.5pt, rounded corners=2pt] at (axis cs:168.5,65.388) {S};
\node[anchor=south west, font=\scriptsize\bfseries, text=MerconCoral, fill=white, fill opacity=0.92, inner sep=1.5pt, rounded corners=2pt] at (axis cs:89.3,65.994) {S};

\draw[MerconSlate!60, dashed, thick] (axis cs:35,65.388) -- (axis cs:680,65.388);
\node[anchor=south east, font=\scriptsize, text=MerconSlate!80!black] at (axis cs:640,65.45) {Best RMSE};

\end{axis}
\end{tikzpicture}
  \vspace{-5pt}
  \caption{Folsom accuracy--throughput trade-off for evaluated strict
           runs. Lower RMSE and higher FPS are preferred.}
  \label{fig:folsom_pareto}
\end{figure}
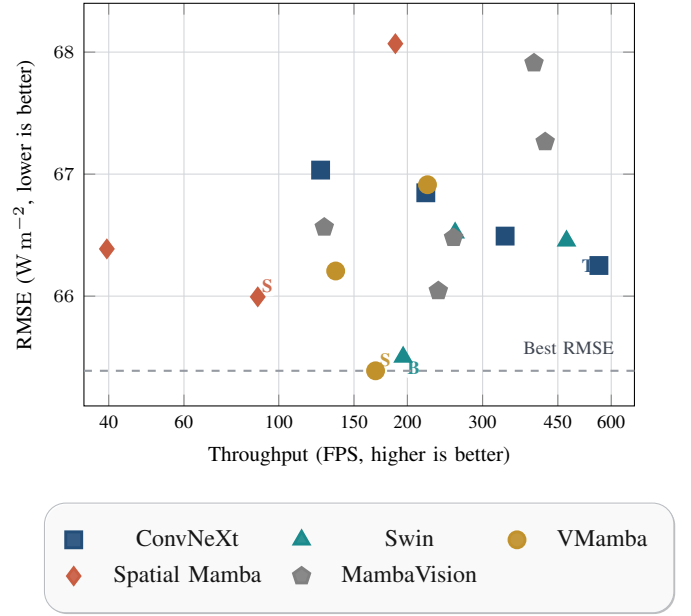

Figure~\ref{fig:folsom_pareto} separates low-error and
high-throughput operating points. VMamba~S gives the lowest Folsom
RMSE, but it is not on the high-throughput edge. ConvNeXt~T and
Swin~T retain most of the Folsom skill at much higher frame rates,
while Spatial Mamba~S is competitive in RMSE but slower than several
alternatives with similar accuracy. The MambaVision rows further
show that parameter count and throughput are not reliable proxies for
forecast quality in this scalar regression setting.

Table~\ref{tab:secondary_metrics} reports secondary metrics for
representative operating points. On Folsom, VMamba~S and Swin~B are
nearly tied in RMSE and $R^{2}$, but Swin~B has lower MAE and
smaller negative bias. Thus, the lowest-RMSE row is not necessarily
the best choice for every operating objective. On NREL, the
temporal-only diagnostic has lower RMSE, lower MAE, and higher
$R^{2}$ than the listed visual rows, reinforcing that visual
capacity alone is not sufficient under this strict low-data split.

\begin{table}[t]
\centering
\caption{Representative secondary metrics from strict test outputs.
         RMSE, MAE, and MBE are in \si{\watt\per\metre\squared}.}
\label{tab:secondary_metrics}
\small
\setlength{\tabcolsep}{4pt}
\renewcommand{\arraystretch}{1.15}
\begin{tabular}{ll S[table-format=2.2]
                    S[table-format=2.2]
                    S[table-format=-1.2]
                    S[table-format=-1.4]}
\toprule
\ColHeader{Site} & \ColHeader{Model} &
{\ColHeader{RMSE}} & {\ColHeader{MAE}} &
{\ColHeader{MBE}}  & {\boldmath$R^{2}$} \\
\midrule
\multirow{3}{*}{Folsom}
  & VMamba S      & 65.39 & 31.61 & -3.38 &  0.9456 \\
  & Swin B        & 65.50 & 30.36 & -2.10 &  0.9454 \\
  & ConvNeXt T    & 66.25 & 31.46 &  2.74 &  0.9441 \\
\midrule
\multirow{3}{*}{NREL}
  & Temporal-only & 21.33 & 12.76 & -3.72 &  0.2501 \\
  & Swin T        & 23.76 & 16.40 & -4.89 &  0.0690 \\
  & MambaVision L & 28.91 & 22.30 &  4.11 & -0.3775 \\
\bottomrule
\end{tabular}
\end{table}

\vspace{-4pt}
\subsection{Effective Receptive Fields}
\begin{figure}[H]
\centering
\setlength{\tabcolsep}{2pt}
\renewcommand{\arraystretch}{0.75}
\begin{tabular}{cc}
    \includegraphics[width=0.40\columnwidth]{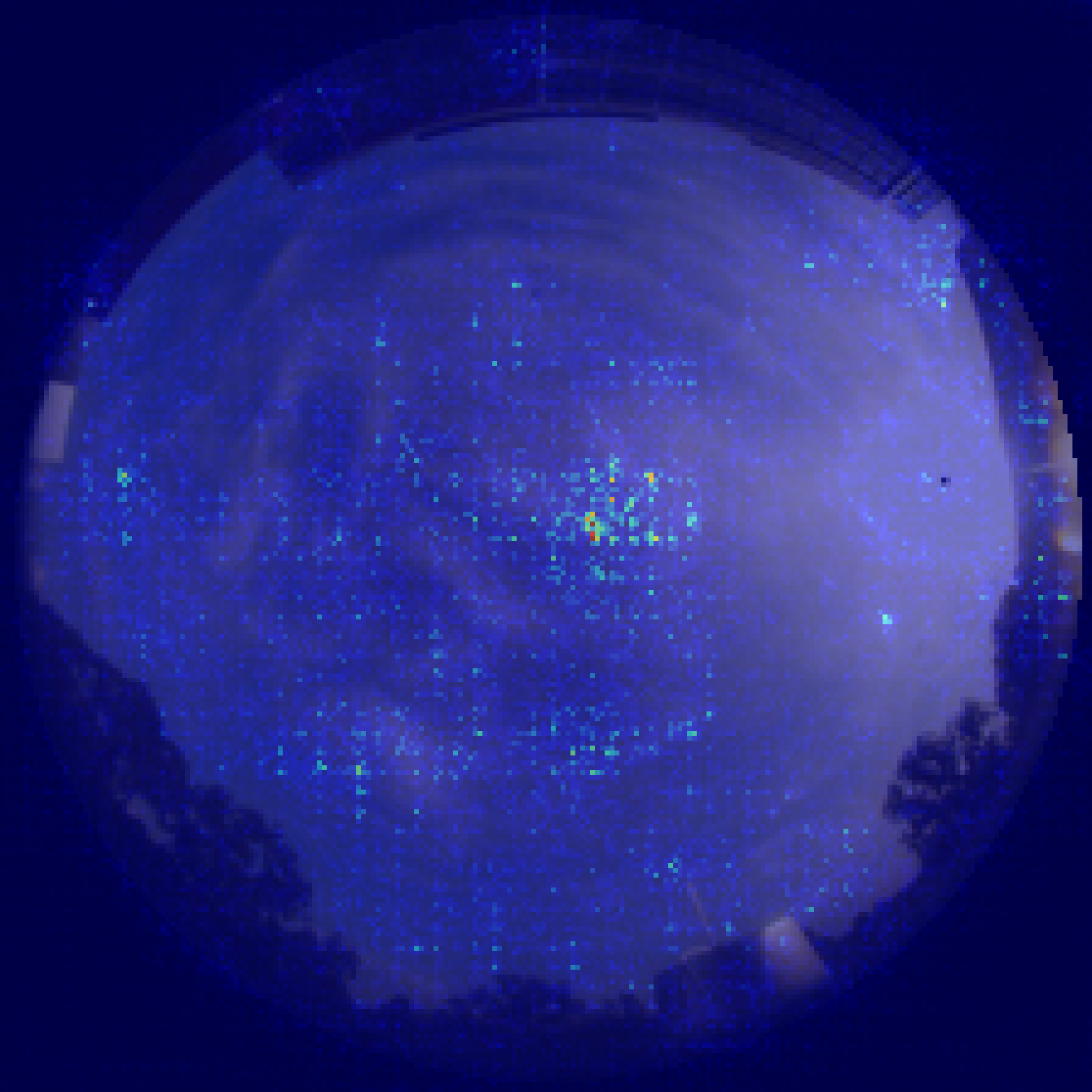} &
    \includegraphics[width=0.40\columnwidth]{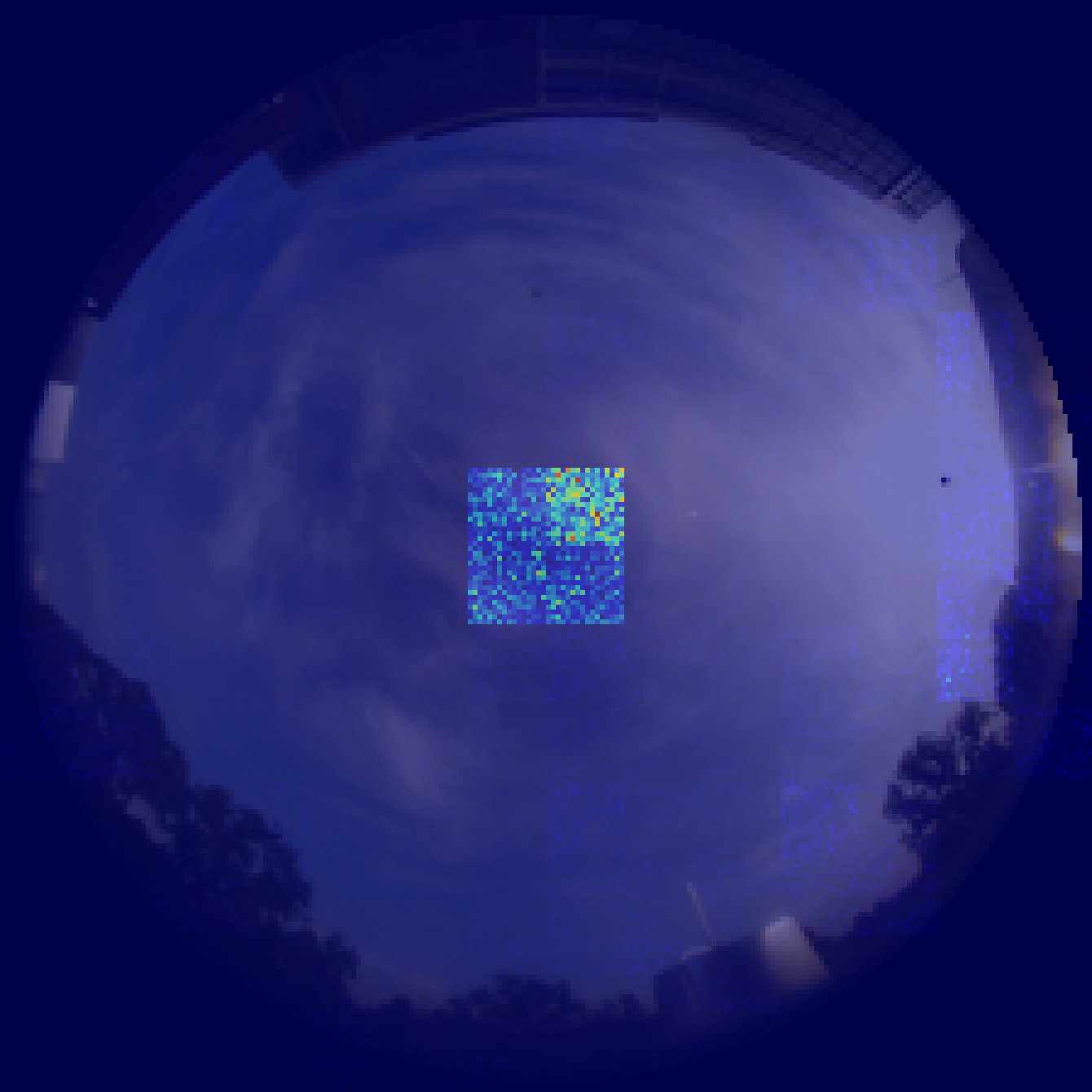} \\
    {\scriptsize ConvNeXt T} &
    {\scriptsize Swin B} \\
    \includegraphics[width=0.40\columnwidth]{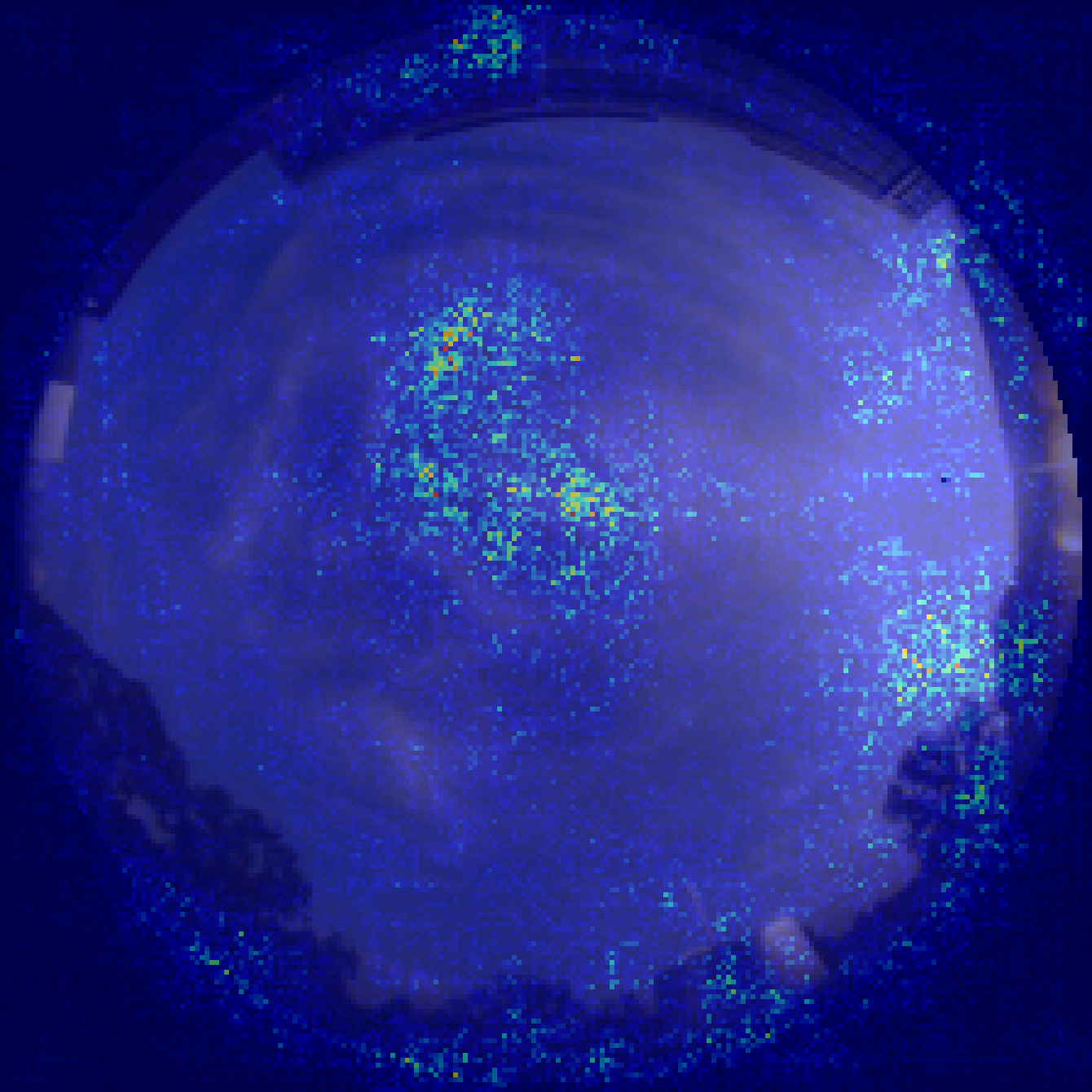} &
    \includegraphics[width=0.40\columnwidth]{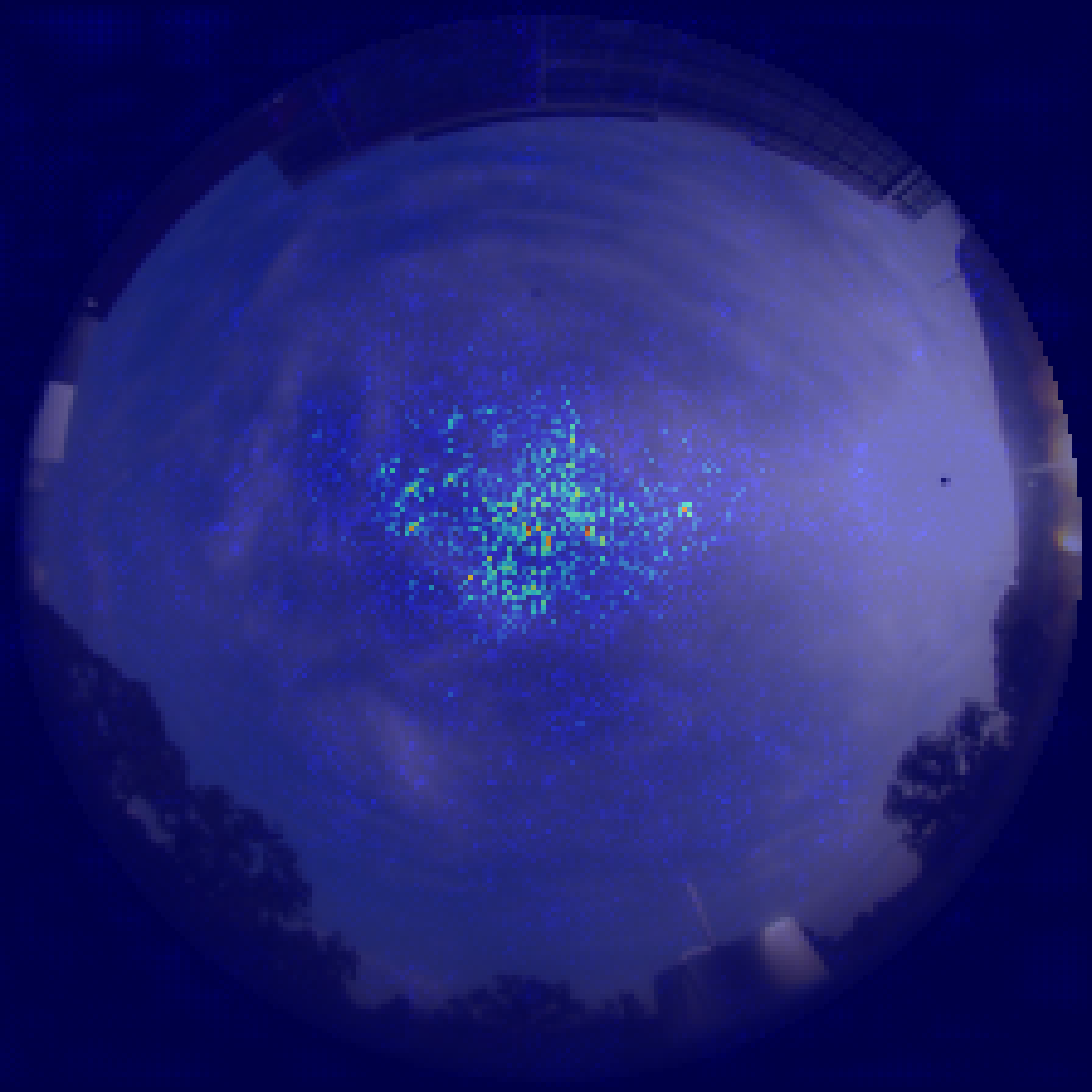} \\
    {\scriptsize VMamba S} &
    {\scriptsize MambaVision T2}
\end{tabular}

\caption{Representative Folsom ERF overlays under the fixed
        forecasting head.}
\label{fig:erf_gallery}
\end{figure}

Fig.~\ref{fig:erf_gallery} provides a qualitative check on how
representative encoders use the sky dome inside the same forecasting
head. ConvNeXt~T shows a diffuse response over much of the dome.
Swin~B is more compact and centred, consistent with localised window
aggregation. VMamba~S spreads sensitivity more broadly and includes
off-centre regions, while MambaVision~T2 lies between the Swin and
VMamba patterns. These overlays do not establish causality; they only
indicate that the backbone families gather spatial context
differently under the fixed head.

\subsection{Scaling and Site Dependence}

Larger encoders do not consistently improve the forecast under the
fixed head. On Folsom, ConvNeXt~L is worse than ConvNeXt~T,
VMamba~B is worse than VMamba~S, and MambaVision~L is worse than
MambaVision~T2. Since the target is a scalar \SI{10}{min}-ahead
clear-sky index, additional visual capacity does not automatically
translate into lower RMSE. This does not show that larger backbones
are intrinsically worse for irradiance forecasting; under the shared
lightweight fusion head and fixed training budget, their additional
capacity may be underused or more prone to overfitting.

The site-level contrast is also important. Folsom has \num{224022}
strict test samples and all evaluated visual-backbone runs beat smart
persistence. NREL has only \num{313} strict test samples after image
matching, daylight filtering, and the \num{40}-step history
requirement; every visual row is worse than smart persistence. A
visual-backbone ranking selected on one site should therefore not be
transferred without checking simple baselines and matched-sample
conditions.

\subsection{Limitations and Conservative Reading}

The Folsom ranking should be read as a controlled single-protocol
comparison, not as a statistically final top-one claim. VMamba~S,
Swin~B, and Spatial Mamba~S are separated by only
\SI{0.60}{\watt\per\metre\squared} RMSE, and VMamba~S and Swin~B
differ by only \SI{0.11}{\watt\per\metre\squared}. Since the top-3
Folsom models span this narrow margin and the efficiency findings are
seed-independent by construction, the qualitative Pareto conclusions
are unlikely to reverse under repeated seeds, though the precise
rank order requires confirmation with repeated-seed evaluation and
paired confidence intervals.

The fixed optimizer schedule, fusion rule, and regression head
improve comparability because only the visual backbone is
intentionally varied. However, the fixed head may not be equally
favourable to all encoder families. The benchmark therefore measures
shared-protocol operating-point behaviour, not each backbone's
individually optimised potential. The temporal-only row is also a
diagnostic for the shared weather-history branch, not a separately
optimised temporal forecasting system.

Finally, the protocol fixes history length in steps rather than
physical duration. On NREL, the \num{40}-step history corresponds to
about \SI{6.5}{h}, contributing to the small matched split. The
highest-value extensions are repeated seeds for the top Folsom
models, paired uncertainty estimates, sky-condition stratification,
cross-site transfer, and limited sweeps of backbone learning rate or
fusion depth.

\section{Conclusion}

We presented a controlled visual-backbone benchmark for multimodal \SI{10}{min}-ahead solar irradiance forecasting. By freezing preprocessing, clear-sky-index target definition, temporal encoding, fusion, regression head, optimizer schedule, seed, and chronological split policy, the study isolates the visual encoder as the only intentionally varied component. The goal is not to propose a fully optimized forecasting system, but to measure how pretrained visual backbones behave inside one common multimodal forecasting protocol.

The evaluated results support three conclusions. First, the Folsom strict split provides a useful visual-encoder comparison: all evaluated visual-backbone runs beat clear-sky-index smart persistence, and VMamba S gives the lowest Folsom visual RMSE. However, Swin B is only \SI{0.11}{\watt\per\metre\squared} worse in RMSE and has lower MAE and smaller bias, so the result should be read as a close single-seed comparison rather than a statistically final winner. Second, larger encoders do not consistently improve scalar irradiance forecasting under the fixed head. Compact ConvNeXt and Swin variants provide much higher throughput with small RMSE penalties, while larger models may be limited by the shared lightweight fusion head and fixed training budget. Third, the \num{313}-sample NREL strict split behaves differently: smart persistence remains strongest, the temporal-only diagnostic is better than the visual rows, and the visual ranking should be interpreted as behavior under extreme matched-sample scarcity rather than as a universal site-level ordering.

These findings suggest that visual-backbone selection for sky-image irradiance forecasting should be reported together with simple baselines, temporal diagnostics, and efficiency measurements. The most important next step is not to add more backbones, but to strengthen the evidence around the current benchmark: repeated-seed evaluation for the top Folsom models, paired confidence intervals over test samples or test days, sky-condition and ramp-event stratification, cross-site transfer, and limited tuning of backbone learning rates and fusion depth. A second direction is to compare the fixed multimodal head with stronger image-only and optimized temporal-only baselines, so that the marginal value of visual information can be measured directly. These extensions would turn the present benchmark into a broader reliability study of visual representation learning for operational solar forecasting.

\FloatBarrier

\bibliographystyle{IEEEtran}
\bibliography{references}

\end{document}